\documentclass[a4paper,12pt]{article}
\begin{document}

\begin{titlepage}
\thispagestyle{empty}

\begin{flushright}
\end{flushright}
\vspace{5mm}

\begin{center}
{\Large \bf
Representations of $p$-brane topological charge algebras}
\end{center}

\begin{center}
{\large D. T. Reimers}\\
\vspace{2mm}

\footnotesize{
{\it School of Physics, The University of Western Australia\\
Crawley, W.A. 6009, Australia}
} \\
{\tt  reimers@physics.uwa.edu.au}\\
\end{center}
\vspace{5mm}

\begin{abstract}
\baselineskip=14pt

The known extended algebras associated with $p$-branes are shown to be generated as topological charge algebras of the standard $p$-brane actions. A representation of the charges in terms of superspace forms is constructed. The charges are shown to be the same in standard/extended superspace formulations of the action.

\end{abstract}

\vfill
\end{titlepage}

\renewcommand{\a}[0]{\alpha}
\renewcommand{\b}[0]{\beta}
\newcommand{\g}[0]{\gamma}
\newcommand{\G}[0]{\Gamma}
\renewcommand{\d}[0]{\delta}
\newcommand{\e}[0]{\epsilon}
\renewcommand{\k}[0]{\kappa}
\newcommand{\m}[0]{\mu}
\newcommand{\n}[0]{\nu}
\newcommand{\p}[0]{\phi}
\renewcommand{\r}[0]{\rho}
\newcommand{\s}[0]{\sigma}
\renewcommand{\S}[0]{\Sigma}
\renewcommand{\t}[0]{\theta}
\newcommand{\x}[0]{\xi}
\newcommand{\D}[0]{\Delta}

\newcommand{\lsb}[0]{\left [}
\newcommand{\rsb}[0]{\right ]}
\newcommand{\lcb}[0]{\left \{}
\newcommand{\rcb}[0]{\right \}}
\newcommand{\blsb}[0]{\bigg [}
\newcommand{\brsb}[0]{\bigg ]}
\newcommand{\half}[0]{\frac{1}{2}}
\newcommand{\third}[0]{\frac{1}{3}}
\newcommand{\quart}[0]{\frac{1}{4}}
\newcommand{\sixth}[0]{\frac{1}{6}}
\newcommand{\sa}[0]{\overrightarrow{\s}}
\newcommand{\nn}[0]{\nonumber}
\newcommand{\slsh}[1]{\overrightarrow{#1}}
\newcommand{\gph}[0]{(-g)^{\half}}
\newcommand{\gmh}[0]{(-g)^{-\half}}
\newcommand{\tb}[0]{\overline\theta}
\newcommand{\del}[0]{\partial}
\renewcommand{\bar}[0]{\overline}
\newcommand{\be}[0]{\begin{equation}}
\newcommand{\ee}[0]{\end{equation}}
\newcommand{\bea}[0]{\begin{eqnarray}}
\newcommand{\eea}[0]{\end{eqnarray}}

\newcommand{\gdl}[1]{\overline{\theta}\Gamma_{#1}d\theta}
\newcommand{\gdu}[1]{\overline{\theta}\Gamma^{#1}d\theta}
\newcommand{\gl}[2]{(\Gamma_{#1} \theta)_{#2}}
\newcommand{\gu}[2]{(\Gamma^{#1} \theta)_{#2}}
\newcommand{\dtgtl}[1]{d\overline{\theta}\Gamma_{#1}\theta}
\newcommand{\dtgtu}[1]{d\overline{\theta}\Gamma^{#1}\theta}
\newcommand{\tgdtl}[1]{\overline{\theta}\Gamma_{#1}d\theta}
\newcommand{\tgdtu}[1]{\overline{\theta}\Gamma^{#1}d\theta}

\section{Introduction}

The $p$-brane Lagrangian \cite{Bergshoeff87,ach87,evans88} consists of the kinetic term and the WZ (Wess-Zumino) term. The field strength of the WZ term has uniqueness and cohomological nontriviality as characteristic properties \cite{azc89-2}. Under the action of the super-Poincar\'{e} group, the $p$-brane Lagrangian is invariant only up to a total derivative that results from the WZ term. Due to this ``quasi-invariance," the Noether charge algebra of the $p$-brane is modified by a topological ``anomalous term" \cite{azc89}. Anomalous term and WZ term are related cohomologically: the former may be found from the latter by solving descent equations in a construction involving ghost fields \cite{azc91,Bergshoeff98}. Based on topological distinctions between bosonic and fermionic coordinates \cite{dewitt}, terms associated with fermionic topology have usually been omitted from anomalous term calculations. This results in bosonic, ``central" extensions of the standard supertranslation algebra (for example, those explicitly derived in \cite{azc89,hammer97}).

On the other hand, there also exist fermionic extensions of the standard supertranslation algebra \cite{green89}. Some of these algebras allow manifestly super-Poincar\'{e} invariant WZ terms to be constructed for the $p$-brane action \cite{siegel94,bergshoeff95,chrys99}. Such extensions (which are in general non-central) contain additional fermionic generators which appear like fermionic analogs of the bosonic topological charges \cite{sezgin95,sezgin96}. The explicit construction of such fermionic topological charges was considered in \cite{hatsuda00,hatsuda01,peeters03}. In the extended superspace formulation of the action the Noether charges associated with extra coordinates are also topological, and a correspondence between the bosonic topological terms in standard/extended formulations of the action was discovered \cite{chrys99}.

Recently we further investigated the cohomological descent system. A total differential approach was established in which the WZ field strength and the anomalous term are \textit{equivalent} representatives of a $(p+2)$-cocycle associated with the $p$-brane \cite{reimers05}. Due to freedom in the choice of representatives, the anomalous term is a cohomology class. The differentials involved in the descent sequence were shown to be equivalent, which implies invertibility of the sequence and that the anomalous term is a unique and nontrivial class. The different representatives of the class result in the generation of a ``spectrum" of topological charge algebras, all of which are extensions of the super-Poincar\'{e} algebra by an ideal. When the terms associated with fermionic topology are retained, one finds that the superspaces underlying extended superspace formulations of the superstring action are generated as topological charge algebras of the standard superstring action.

The main purpose of this paper is to show that this correspondence continues for $p$-branes with $p\geq 2$. Since the results of \cite{azc89} exclude not only fermionic charges, but also the fermionic corrections to the bosonic charges, the generalization of these results where all terms are retained is required (the simplifications associated with trivial fermionic topology may be deduced at the end). We find this generalization, not by the descent method but by using uniqueness of the anomalous term. The charges are shown to be representations of the ideals of the extended algebras of \cite{bergshoeff95,chrys99}. It follows that these extended algebras are indeed generated as topological charge algebras of the standard $p$-brane action. It emerges along the way that the topological charges are the unique solution satisfying the extended algebra, and that the charges (including \textit{all} terms both bosonic and fermionic) are the same in standard/extended formulations of the action.

The structure of this paper is as follows. In section \ref{sec:Preliminaries}, our notation is introduced and the properties of $p$-branes are summarized. The construction of topological charge algebras is reviewed and a summary of the descent methods is given. In section \ref{sec:Anomalous term representatives}, we present the closed forms that provide representations of the ideals of the known extended algebras associated with $p$-branes. An associated form is shown to be a representative of the anomalous term of the Noether charge algebra of the standard superspace $p$-brane action. In section \ref{sec:extended Noether charges}, it is shown that the derived forms also represent Noether charges for $p$-branes defined on the corresponding extended superspaces. In section \ref{sec:Comments}, we comment on some properties of the results.

\section{Preliminaries}
\label{sec:Preliminaries}

\subsection{$p$-branes}

We start with a brief review of the required supergroup equations. Useful references on this material include \cite{Zumino:1977av,green89,azc89-2,azc91,bergshoeff95,chrys99}, with more comprehensive treatments in \cite{dewitt,kuzenko-book}. The superalgebra of the supertranslation group is\footnote{The charge conjugation matrix will not be explicitly shown. It will only be used to raise/lower indices on gamma matrices, which have the standard position $\Gamma^{\a}{}_{\b}$. $\Gamma_{\a\b}$ is assumed to be symmetric. Majorana spinors are assumed throughout (thus, for example, $\bar\t_{\a}=\t^{\b}C_{\b\a}$). The right acting convention for the de Rham differential is used, and wedge product multiplication of forms is understood.}:
\be
    \lcb Q_{\alpha},Q_{\beta}\rcb =\Gamma^{a}{}_{\alpha\beta}P_{a}.
\ee
The corresponding group manifold can be parameterized:
\bea
    g(Z)&=&e^{x^{a}P_{a}}e^{\theta^{\a}Q_{\a}}\\
    Z^{A}&=&(x^{a},\theta^{\a}).\nn
\eea
The left vielbein is defined by:
\bea
    L(Z)&=&g^{-1}(Z)dg(Z)\\
    &=&dZ^{M}L_{M}{}^{A}(Z)T_{A}\nn,
\eea
where $T_A$ represents the full set of superalgebra generators. The right vielbein is defined similarly:
\bea
    R(Z)&=&dg(Z)g^{-1}(Z)\\
    &=&dZ^{M}R_{M}{}^{A}(Z)T_{A}\nn.
\eea
The left group action is defined by:
\be
    g(Z')=g(\e)g(Z),
\ee
where $\e^A$ is an infinitesimal constant. The corresponding superspace transformation is generated by the operators:
\be
    \label{left generators}
     Q_{A}=R_{A}{}^{M}\del_{M},
\ee
where $R_{A}{}^{M}$ are the inverse right vielbein components, defined by:
\be
    R_{A}{}^{M}R_{M}{}^{B}=\d_{A}{}^{B}.
\ee
Explicitly this yields:
\be
\begin{array}{lll}
    &Q_\a x^m=-\half (\G^m\t)_\a,\qquad &Q_\a \t^\m=\d_\a{}^\m,\\
    &Q_a x^m=\d_a{}^m,\qquad &Q_a \t^\m=0\nn.
\end{array}
\ee
$Q_A$ are the generators of the left group action, and will be referred to as the ``left generators." The action of $Q_{A}$ upon superspace forms is given by the Lie derivative with respect to the vector field associated with (\ref{left generators}). Forms that are invariant under the global left group action will be called ``left invariant." The vielbein components $L^{A}$ are left invariant by construction. Their explicit form is:
\begin{eqnarray}
    L^{a}&=&dx^{a}-\half \dtgtu{a}\\
    L^{\alpha}&=&d\theta^{\alpha}\nn.
\end{eqnarray}
Indices $A,B,C,D$ will be used to indicate components with respect to this basis. Indices $M,N,L,P$ will be used for the coordinate basis.

The NG (Nambu-Goto) action for a $p+1$ dimensional manifold embedded in the background superspace is:
\be
    S=-\int d^{p+1}\s\sqrt{-g}.
\ee
The integral is over the embedded $p+1$ dimensional ``worldvolume," which has coordinates $\s^{i}$. The worldvolume metric $g_{ij}$ is defined using the pullback of the left vielbein:
\bea
    L_{i}{}^{A}&=&\del_{i}Z^{M}L_{M}{}^{A}\\
    g_{ij}&=&L_{i}{}^{a}L_{j}{}^{b}\eta_{ab}\nn,
\eea
and $g$ denotes $\det g_{ij}$. A $p$-brane is the $\k$-symmetric generalization of the NG action. The $p$-brane action is \cite{Bergshoeff87,ach87,evans88}:
\be
    \label{action}
    S=-\int d^{p+1}\s\sqrt{-g}+\int B.
\ee
The first term is the ``kinetic" term. The second term is the WZ term, which is the integral over the worldvolume of a superspace form $B$ defined by the property \cite{Bergshoeff87,ach87}:
\bea
    \label{H def}
    dB&=&H\\
    &\propto& d\t^{\a}d\t^{\b}L^{a_{1}}\ldots L^{a_{p}}(\G_{a_{1}\ldots a_{p}})_{\a\b}.\nn
\eea
The proportionality constant depends on $p$ and is determined by requiring $\k$-symmetry of the action. Closure of $H$ requires the Fierz identity \cite{Bergshoeff87,ach87,evans88}:
\bea
    \G^{[a_{1}\ldots a_{p}]}{}_{(\a\b}\G_{a_p\d\e)}&=&0,
\eea
which is only satisfied for certain combinations of $p$ (number of spatial brane dimensions) and $d$ (superspace dimension). The allowed values of $(p,d)$ (called the ``minimal branescan") are such that:
\be
    (\G_{[a_{1}\ldots a_{p}]})_{\a\b}=(\G_{[a_{1}\ldots a_{p}]})_{\b\a}.
\ee
This ensures that $H$ can be nonzero. It turns out that $H$ is the unique, closed, left invariant $(p+2)$-form of dimension $p+1$ \cite{azc89-2}.

\subsection{Topological charge algebras}
\label{sec:Modified charge algebras}

We are familiar with Noether charge algebras in which symmetries of an action are associated with conserved charges that transform according to the underlying symmetry group. Topological extensions to supersymmetry algebras can occur if the topology is such that surface terms contribute to the charge algebra \cite{witten78}. The topological charge algebras considered here are those which generalizes the Noether construction to the case of actions which are invariant only up to a total derivative; the case with $p$-branes \cite{azc89}. A quite general treatment of this material is given in \cite{hammer97}. We now give a brief review.

The Hamiltonian formulation of dynamics is cast in terms of coordinates $Z^{M}$ and their associated conjugate momenta $P_{M}$, which together constitute the ``phase space." The momenta are defined by:
\be
    P_{M}=\frac{\del L}{\del \dot Z^{M}}.
\ee
The following fundamental (graded) Poisson brackets on phase space will be used\footnote{Different types of bracket operation are used in this paper. We will not explicitly indicate the type since this should be clear within context.}:
\be
    \lsb P_{M}(\s),Z^{N}(\s')\right \}=\d_{M}{}^{N}\d(\sa-\sa'),
\ee
where it is assumed $\s'^{0}=\s^{0}$ (i.e. equal time brackets). The Dirac delta function notation is shorthand for the product of the $p$ delta functions associated with the spatial coordinates of the worldvolume.

The Noether charges associated with a manifestly left invariant Lagrangian will be denoted $\bar Q_A$. One finds:
\be
    \bar Q_{A}=\int d^{p}\s R_{A}{}^{M}P_{M}.
\ee
These charges satisfy the same algebra as the underlying superalgebra, but with the sign reversed:
\be
    \label{minimal algebra}
    \lsb \bar Q_{A},\bar Q_{B}\rcb =-t_{AB}{}^{C}\bar Q_{C},
\ee
where $t_{AB}{}^C$ are the structure constants of the underlying superalgebra. For later convenience we refer to (\ref{minimal algebra}) as the ``minimal algebra." It follows from the left invariance of $H$ that the left variation of the WZ form $B$ is closed \cite{Bergshoeff87,ach87,evans88,azc89-2}:
\be
    \label{Q_A=-dW_A}
    Q_AB=-dW_A.
\ee
If $Q_AB\neq 0$, the $p$-brane Lagrangian is symmetric only up to a total derivative. We define a ``bar map" by its action on an arbitrary superspace $p$-form $Y$:
\be
    \label{bar map}
    \bar Y=(-1)^p\int \Phi^* Y.
\ee
The map $\Phi$ embeds the spatial section of the worldvolume, which we assume to be a closed manifold. Due to the variation (\ref{Q_A=-dW_A}), the conserved charges in the presence of the WZ term are \cite{azc89,hammer97}:
\be
	\label{conserved charges}
    \widetilde {\bar Q}_{A}=\bar Q_{A}+\bar W_{A}.
\ee
The conserved charges obey a modified version of the minimal algebra \cite{azc89,hammer97}:
\be
	\label{algebra of conserved charges}
    \lsb \widetilde {\bar Q}_{A},\widetilde {\bar Q}_{B}\rcb =-t_{AB}{}^{C}\widetilde {\bar Q}_{C}+\bar M_{AB},
\ee
with
\be
    \label{topological anom term}
    \bar M_{AB}=\lsb \bar Q_{A},\bar W_{B}\rcb +\lsb \bar W_{A},\bar Q_{B}\rcb +t_{AB}{}^{C}\bar W_{C}.
\ee
$\bar M$ is the topological ``anomalous term" which modifies the Noether charge algebra.

\subsection{Anomalous term cohomology}

The de Rham complex consists of the space of differential forms under the action of the exterior derivative $d$. This can be extended into a double complex by the addition of a second nilpotent operator that commutes with $d$ (see \cite{bott82} for a comprehensive treatment). The operator used in this paper is a ``ghost differential" $s$ which requires the introduction of a ghost partner $e^{A}$ for each coordinate \cite{azc91}. The ghost fields have the opposite grading to coordinates:
\bea
    \lsb e^{A},Z^{M}\rcb &=&0\\
    \lcb e^{A},e^{B}\rsb &=&0\nn,
\eea
where $[\quad,\quad \}$ and $\{\quad ,\quad ]$ are the graded commutator/anticommutator. The $e^A$ are independent of the coordinates $Z^{M}$ and hence satisfy $de^{A}=0$. A general element of the double complex is a ``ghost form valued differential form." The space of all such ``generalized forms" of differential degree $m$ and ghost degree $n$ will be denoted by $\Omega^{m,n}$. A generalized form $Y\in \Omega^{m,n}$ will be written using a comma to separate ghost indices from space indices:
\be
    Y=e^{B_{n}}\ldots e^{B_{1}}L^{A_{m}}\ldots L^{A_{1}}Y_{A_{1}\ldots A_{m},B_{1}\ldots B_{n}}\frac{1}{m!n!}.
\ee
The ghost differential can be defined by the properties:
\begin{itemize}
\item
$s$ is a right derivation. That is, if $X$ and $Y$ are generalized forms and $n$ is the ghost degree of $Y$ then:
\be
    s(XY)=Xs(Y)+(-1)^{n}s(X)Y.
\ee
\item
If $X$ has ghost degree zero then:
\be
    sX=e^{A}Q_{A}X.
\ee
\item
\be
    se^{A}=\half e^{C}e^{B}t_{BC}{}^{A}.
\ee
\end{itemize}

There is a total differential $D$ that is naturally associated with a double complex \cite{bott82}, which in this case is \cite{reimers05}:
\bea
    D&=&s+(-1)^{n+1}d\\
    D^{2}&=&0\nn,
\eea
where $n$ is the ghost degree of the generalized form upon which $D$ acts. The spaces $\Omega^l_{D}$ of the single complex upon which $D$ acts are the sum along the anti-diagonal of the spaces of the double complex:
\be
    \Omega_{D}^{l}=\{ \oplus\Omega^{m,n}:\quad m+n=l \} .
\ee
The $l$-th cohomology of $D$ is:
\be
    H_D^{l}=Z_D^{l}/B_D^{l},
\ee
where $Z_D^{l}$ are the $D$ closed generalized $l$-forms (``$D$ cocycles"), and $B_D^{l}$ are the generalized $l$-forms in the image of $D$ (``$D$ coboundaries"). The restriction of $H_D^l$ to representatives within $\Omega^{m,l-m}$ will be denoted $H^{m,l-m}$.

The $p$-brane has an associated $D$ cocycle defined by the representative $H\in H^{p+2,0}$, with $H$ as given in (\ref{H def}). One progresses from $H$ to the anomalous term via ``descent equations" \cite{azc91}. The first two descent equations are \cite{azc91,Bergshoeff98,reimers05}:
\bea
    H&=&dB\\
    sB&=&-dW\nn.
\eea
The anomalous term can then be represented by the form \cite{azc91,Bergshoeff98,reimers05}:
\be
    M=sW,
\ee
which is the $H^{p,2}$ representative for the $D$ cocycle. The topological anomalous term (\ref{topological anom term}) is related to this via the map (\ref{bar map}). Because $M$ is $d$ closed, $\bar M$ is a topological integral of $M$ over the spatial section of the worldvolume.

It's well known that equation (\ref{H def}) defines $B$ only up to a total derivative. In the cocycle description this is part of the gauge freedom generated by $D$ coboundaries. The transformations for $B$ and $W$ are generated by gauge fields $\psi\in\Omega^{p,0}$ and $\lambda\in\Omega^{p-1,1}$ \cite{reimers05}:
\bea
    \D B&=&-d\psi\\
    \D W&=&s\psi+d\lambda\nn.
\eea
The resulting gauge transformation of the anomalous term is:
\be
	\label{M freedom}
    \D M=sd\lambda.
\ee
All elements of the double complex (including the gauge fields) must satisfy the requirements of Lorentz invariance and dimensionality $p+1$.

Now $H$ is the unique Poincar\'{e} invariant, $d$ closed form of dimensionality $p+1$ \cite{azc89-2} (uniqueness is up to a proportionality constant). As a result there are no coboundaries for the $H^{p+2,0}$ cohomology. However, there \textit{are} coboundaries for the $H^{p,2}$ cohomology; this is the gauge freedom (\ref{M freedom}) for $M$. So the anomalous term is well defined only as the cohomology class $[M]$ consisting of the restriction of $H^{p,2}$ to Lorentz invariant forms of dimensionality $p+1$. This class is nontrivial and unique \cite{reimers05}. As a result, if we can find a single nontrivial representative for $[M]$, the entire class will be generated by the $\lambda$ gauge transformations.

As in \cite{Bergshoeff98}, we find it easiest to work with differential operators and the forms from which the Noether charges derive instead of the Noether charges themselves. Instead of (\ref{conserved charges}) we thus use left generators modified by forms \cite{azc91,Bergshoeff98,reimers05}:
\be
    \label{modified left generators}
    \widetilde Q_A=Q_A+W_A.
\ee
These obey the same modified algebra (\ref{algebra of conserved charges}) as the conserved charges \cite{azc91,Bergshoeff98,reimers05}:
\be
    \lsb \widetilde {Q}_{A},\widetilde {Q}_{B}\rcb =-t_{AB}{}^{C}\widetilde {Q}_{C}+M_{AB}.
\ee
The full class $[M]$ therefore generates a ``spectrum" of extended superalgebras. If the fermionic topology is trivial, $M$ generates bosonic, ``central" extensions of the supertranslation group \cite{azc89}. In the general case, the representatives $M$ continue to generate extensions of the standard supertranslation algebra, but the extensions are now in general fermionic and non-central \cite{reimers05}. These ``operator-form" representations of the algebras contain operators $\widetilde Q_A$, and extra generators represented by closed superspace forms $\S_{\check A}$. The associated topological charge algebra (\ref{algebra of conserved charges}) is obtained by the replacement:
\bea
    \widetilde {Q}_A&\rightarrow &\widetilde {\bar Q}_A\\
    \S_{\check A}&\rightarrow &\bar \S_{\check A}\nn.
\eea

\section{$p$-brane topological charge algebras}
\label{sec:Anomalous term representatives}

For higher values of $p$, finding the anomalous term via descent equations becomes lengthy. In this paper we will make use of the uniqueness of the anomalous term instead. We wish to find a Lorentz invariant, $D$ nontrivial element:
\be
    M^{(p)}\in H^{p,2}
\ee
of dimensionality $p+1$, for each allowed value of $p$. By uniqueness of the class, this must then be a representative of the $p$-brane anomalous term. If required, the full class $[M]$ can be generated by applying the $\lambda$ gauge transformations to this representative. There is no a priori obvious way to find $M^{(p)}$. However, we are motivated by the observation that the spectrum of topological charge algebras of the string action \cite{reimers05} consisted of extended superalgebras that allow left invariant WZ forms to be constructed for the string action. This spectrum contained three different types of algebra (when classified according to the generators present). Two of these algebras had been previously used to construct invariant actions: the Green algebra \cite{green89} used in \cite{siegel94}, and also a four generator extension \cite{bergshoeff95,chrys99}. An algebra which allows a left invariant WZ form to be constructed for each $p$-brane of higher dimension is also already known. The cases $p=2,3$ were given in \cite{bergshoeff95}. In \cite{chrys99}, an ansatz was presented to generate Maurer-Cartan equations for the required algebra for general values of $p$; however the minimal branescan dictates that $p$-branes exist only for $p\leq 5$ \cite{ach87,evans88}.

In this paper, the approach we will take to find $M^{(p)}$ somewhat reverses the process used in \cite{reimers05}. We begin with the known extended algebra associated with a given value of $p$. We assume that this extended algebra is contained in the spectrum of topological charge algebras generated by the standard superspace $p$-brane action. If this assumption is correct then the extended algebra must have an operator-form representation where the generators of the ideal are represented by closed superspace forms. We will explicitly find these forms. A particular $(p,2)$-form $M^{(p)}$ constructed from them will then be shown to be a representative of the anomalous term associated with the standard superspace $p$-brane action.

For reference, let us give the known extended algebras that allow left invariant WZ terms to be constructed. The algebras will be given in the operator-form convention for which we seek the representation (generators are negatives of those in the corresponding superalgebra underlying the extended superspace action).

\subsection{$p=1$ superalgebra}
\cite{bergshoeff95,chrys99}:
\bea
\label{p=1 algebra}
    \lcb \widetilde Q_{\alpha},\widetilde Q_{\beta}\rcb &=&-\Gamma^{a}{}_{\alpha\beta}\widetilde P_{a}-\Gamma_{a\a\b}\S^{a}\\
    \lsb \widetilde Q_{\a},\widetilde P_{a}\rsb &=&-\Gamma_{a\a\b}\Sigma^{\b}\nn\\
    \lsb \widetilde Q_{\a},\S^a\rsb &=&-\Gamma^a{}_{\a\b}\Sigma^{\b}\nn.
\eea

\subsection{$p=2$ superalgebra}
\cite{bergshoeff95}:
\bea
\label{p=2 algebra}
    \lcb \widetilde Q_\a,\widetilde Q_\b\rcb &=&-\G^a{}_{\a\b}\widetilde P_a-\G_{ab\a\b}\S^{ab}\\
    \lsb \widetilde Q_\a,\widetilde P_a\rsb &=&-\G_{ab\a\b}\S^{b\b}\nn\\
    \lsb \widetilde P_a,\widetilde P_b\rsb &=&-\G_{ab\a\b}\S^{\a\b}\nn\\
    \lsb \widetilde Q_\a,\S^{ab}\rsb &=&-\G^{[a}{}_{\a\b}\S^{b]\b}\nn\\
    \lsb \widetilde P_a,\S^{bc}\rsb &=&-\half\d_a^{[b}\G^{c]}{}_{\a\b}\S^{\a\b}\nn\\
    \lcb \widetilde Q_\a,\S^{a\b}\rcb &=&-\quart\G^{a}{}_{\g\d}\S^{\g\d}\d_\a^\b-2\G^{a}{}_{\a\g}\S^{\g\b}\nn.
\eea

\subsection{$p=3$ superalgebra}
\cite{bergshoeff95}:
\bea
    \lcb \widetilde Q_\a,\widetilde Q_\b\rcb &=&-\G^a{}_{\a\b}\widetilde P_a-\G_{abc\a\b}\S^{abc}\\
    \lsb \widetilde Q_\a,\widetilde P_a\rsb &=&-\G_{abc\a\b}\S^{bc\b}\nn\\
    \lsb \widetilde P_a,\widetilde P_b\rsb &=&-\G_{abc\a\b}\S^{c\a\b}\nn\\
    \lsb \widetilde Q_\a,\S^{abc}\rsb &=&-\G^{[a}{}_{\a\b}\S^{bc]\b}\nn\\
    \lsb \widetilde P_a,\S^{bcd}\rsb &=&-\half\d_a^{[b}\G^{c}{}_{\a\b}\S^{d]\a\b}\nn\\
    \lcb \widetilde Q_\a,\S^{ab\b}\rcb &=&-\quart\G^{[a}{}_{\g\d}\S^{b]\g\d}\d_\a^\b-2\G^{[a}{}_{\a\g}\S^{b]\g\b}\nn\\
    \lsb \widetilde P_a,\S^{bc\a}\rsb &=&-\d_a^{[b}\G^{c]}{}_{\b\g}\S^{\b\g\a}\nn\\
    \lsb \widetilde Q_\a,\S^{a\b\g}\rsb &=&-\half\G^{a}{}_{\d\e}\S^{\d\e(\b}\d_\a^{\g)}-\frac{5}{2}\G^{a}{}_{\a\d}\S^{\d\b\g}\nn.
\eea

\subsection{$p=4$ superalgebra}
Derived from an ansatz for Maurer-Cartan equations in \cite{chrys99}:
\bea
    \lcb \widetilde Q_\a,\widetilde Q_\b\rcb &=&-\G^a{}_{\a\b}\widetilde P_a-\G_{abcd\a\b}\S^{abcd}\\
    \lsb \widetilde Q_\a,\widetilde P_a\rsb &=&-\G_{abcd\a\b}\S^{bcd\b}\nn\\
    \lsb \widetilde P_a,\widetilde P_b\rsb &=&-\G_{abcd\a\b}\S^{cd\a\b}\nn\\
    \lsb \widetilde Q_\a,\S^{abcd}\rsb &=&-\G^{[a}{}_{\a\b}\S^{bcd]\b}\nn\\
    \lsb \widetilde P_a,\S^{bcde}\rsb &=&-\half\d_a^{[b}\G^{c}{}_{\a\b}\S^{de]\a\b}\nn\\
    \lcb \widetilde Q_\a,\S^{abc\b}\rcb &=&-\quart\G^{[a}{}_{\g\d}\S^{bc]\g\d}\d_\a^\b-2\G^{[a}{}_{\a\g}\S^{bc]\g\b}\nn\\
    \lsb \widetilde P_a,\S^{bcd\a}\rsb &=&-\d_a^{[b}\G^{c}{}_{\b\g}\S^{d]\b\g\a}\nn\\
    \lsb \widetilde Q_\a,\S^{ab\b\g}\rsb &=&-\half\G^{[a}{}_{\d\e}\S^{b]\d\e(\b}\d_\a^{\g)}-\frac{5}{2}\G^{[a}{}_{\a\d}\S^{b]\d\b\g}\nn\\
    \lsb \widetilde P_a,\S^{bc\a\b}\rsb &=&-\d_a^{[b}\G^{c]}{}_{\g\d}\S^{\g\d\a\b}\nn\\
    \lcb \widetilde Q_\a,\S^{a\b\g\d}\rcb &=&-\frac{3}{5}\G^{a}{}_{\e\s}\S^{\e\s(\b\g}\d_\a^{\d)}-\frac{12}{5}\G^{a}{}_{\a\e}\S^{\e\b\g\d}\nn.
\eea

\subsection{$p=5$ superalgebra}
Derived from an ansatz for Maurer-Cartan equations in \cite{chrys99}:
\bea
\label{p=5 algebra}
    \lcb \widetilde Q_\a,\widetilde Q_\b\rcb &=&-\G^a{}_{\a\b}\widetilde P_a-\G_{abcde\a\b}\S^{abcde}\\
    \lsb \widetilde Q_\a,\widetilde P_a\rsb &=&-\G_{abcde\a\b}\S^{bcde\b}\nn\\
    \lsb \widetilde P_a,\widetilde P_b\rsb &=&-\G_{abcde\a\b}\S^{cde\a\b}\nn\\
    \lsb \widetilde Q_\a,\S^{abcde}\rsb &=&-\G^{[a}{}_{\a\b}\S^{bcde]\b}\nn\\
    \lsb \widetilde P_a,\S^{bcdef}\rsb &=&-\half\d_a^{[b}\G^{c}{}_{\a\b}\S^{def]\a\b}\nn\\
    \lcb \widetilde Q_\a,\S^{abcd\b}\rcb &=&-\quart\G^{[a}{}_{\g\d}\S^{bcd]\g\d}\d_\a^\b-2\G^{[a}{}_{\a\g}\S^{bcd]\g\b}\nn\\
    \lsb \widetilde P_a,\S^{bcde\a}\rsb &=&-\d_a^{[b}\G^{c}{}_{\b\g}\S^{de]\b\g\a}\nn\\
    \lsb \widetilde Q_\a,\S^{abc\b\g}\rsb &=&-\half\G^{[a}{}_{\d\e}\S^{bc]\d\e(\b}\d_\a^{\g)}-\frac{5}{2}\G^{[a}{}_{\a\d}\S^{bc]\d\b\g}\nn\\
    \lsb \widetilde P_a,\S^{bcd\a\b}\rsb &=&-\d_a^{[b}\G^{c}{}_{\g\d}\S^{d]\g\d\a\b}\nn\\
    \lcb \widetilde Q_\a,\S^{ab\b\g\d}\rcb &=&-\frac{3}{5}\G^{[a}{}_{\e\s}\S^{b]\e\s(\b\g}\d_\a^{\d)}-\frac{12}{5}\G^{[a}{}_{\a\e}\S^{b]\e\b\g\d}\nn\\
    \lsb \widetilde P_a,\S^{bc\a\b\g}\rsb &=&-\d_a^{[b}\G^{c]}{}_{\d\e}\S^{\d\e\a\b\g}\nn\\
    \lsb \widetilde Q_\a,\S^{a\b\g\d\e}\rsb &=&-\frac{5}{6}\G^{a}{}_{\s\r}\S^{\s\r(\b\g\d}\d_\a^{\e)}-\frac{35}{12}\G^{a}{}_{\a\s}\S^{\s\b\g\d\e}\nn.
\eea

We wish to find closed forms $\S^{A_1\ldots A_p}$ satisfying these algebras under the action of the modified left generators (\ref{modified left generators}). If we can, then each extended algebra can be interpreted as the minimal algebra (\ref{minimal algebra}) modified by an anomalous term $M^{(p)}$. The components $M^{(p)}{}_{AB}$ are read as the modifications to the $\lsb Q_A,Q_B\rcb $ brackets of the minimal algebra. For example, from:
\be
    \left \{\widetilde Q_\a,\widetilde Q_\b\right \}=-\G^a{}_{\a\b}\widetilde P_a-\G_{a_1\ldots a_p\a\b}\S^{a_1\ldots a_p}
\ee
we learn that:
\be
    M^{(p)}{}_{\a\b}=-\G_{a_1\ldots a_p\a\b}\S^{a_1\ldots a_p}.
\ee
Reading similarly from the RHS of $\lsb\widetilde Q_\a,\widetilde P_b\rsb$ and $\lsb\widetilde P_a,\widetilde P_b\rsb$, it follows that $M^{(p)}$ has the structure:
\begin{itemize}
\item
$p=1$
\bea
\label{4:p=1 anomalous term}
    M^{(1)}&=&-\half e^\b e^\a \G_{a\a\b}\S^a\\
    &&-e^a e^\a \G_{a\a\b}\S^\b\nn.
\eea
\item
$p\geq 2$
\bea
\label{p greater than 1 anomalous term}
    M^{(p)}&=&-\half e^\b e^\a \G_{a_1\ldots a_p\a\b}\S^{a_1\ldots a_p}\\
    &&-e^a e^\a \G_{aa_1\ldots a_{p-1}\a\b}\S^{a_1\ldots a_{p-1}\b}\nn\\
    &&-\half e^b e^a \G_{aba_1\ldots a_{p-2}\a\b}\S^{a_1\ldots a_{p-2}\a\b}\nn.
\eea
\end{itemize}
To find the required closed forms $\S^{A_1\ldots A_p}$, one firstly observes that:
\be
    \ [\widetilde Q_A,\S^{A_1\ldots A_p}\}=[Q_A,\S^{A_1\ldots A_p}\}.
\ee
The unmodified left generators are thus sufficient for our purposes and the explicit form of $\widetilde Q_A$ is not required. Secondly, the $\S^{A_1\ldots A_p}$ must all have their ``natural" dimension:
\be
    \dim \lsb \S^{a_1\ldots a_m \a_1\ldots \a_n}\rsb =m+\frac{n}{2}.
\ee
This follows from the requirement $\dim M^{(p)}=p+1$, and the fact that $Q_A$ reduces the dimension of a form by the dimension associated with its index. One finally notes that the generator $\S^{\a_1\ldots \a_p}$ is ``central." There is only one candidate for $\S^{\a_1\ldots \a_p}$ satisfying the required properties:
\be
    \S^{\a_1\ldots \a_p}\propto d\t^{\a_1}\ldots d\t^{\a_p}.
\ee
We shall fix the proportionality constant at unity since it serves only as an overall scaling for the extra generators. To find the remaining generators, one can first write the most general allowed form for $\S^{a\a_1\ldots \a_{p-1}}$ using arbitrary coefficients. The coefficients are then found by requiring that the extended superalgebra be satisfied. The process is then continued for $\S^{ab\a_1\ldots \a_{p-2}}$ and so on until the final generator $\S^{a_1\ldots a_p}$ is found. The relevant Fierz identity is required to find the solutions, and its implementation is sometimes more nontrivial than usual due to double symmetrizations which overlap only partially. In general, one finds that the requirement of satisfying the extended superalgebra results in more equations than coefficients present. A solution for such a system is only possible if a sufficient number of the equations are redundant. In fact, exactly the right number of redundant equations are present in order that the solution be unique. That is, the representation for each algebra is \textit{unique}. Having obtained the solution, the redundant equations then provide a good consistency check. We note here that the $p=1,2$ expressions below were also found in \cite{chrys99} in the context of extended superspace actions; more on this in section \ref{sec:extended Noether charges}. The results are\footnote{We anticipate the final result by referring to the forms of the representation as ``charges."}:

\subsection{$p=1$ charges}
\bea
\label{3:p=1 representation}
    \S^{\a}&=&d\t^\a\\
    \S^{a}&=&2dx^a\nn.
\eea
\subsection{$p=2$ charges}
\bea
\label{3:p=2 representation}
    \S^{\a\b}&=&d\blsb d\t^\a \t^\b\brsb\\
    \S^{a\b}&=&d\blsb\frac{9}{2}dx^a\t^\b+\quart\gdu{a}\t^\b\brsb\nn\\
    \S^{ab}&=&d\blsb 5x^adx^b+\half x^{[a}\gdu{b]}\brsb\nn.
\eea
\subsection{$p=3$ charges}
\bea
    \S^{\a\b\g}&=&d\blsb d\t^\a d\t^\b \t^\g\brsb\\
    \S^{a\b\g}&=&d\blsb 6dx^a d\t^\b\t^\g+\half\gdu{a}d\t^{(\b}\t^{\g)}\brsb\nn\\
    \S^{ab\b}&=&d\blsb-\frac{29}{2}dx^a dx^b \t^\b -\frac{3}{2}dx^{[a}\gdu{b]}\t^\b -x^{[a}\gdu{b]}d\t^\b\nn\\
        &&-\frac{1}{8}\gdu{a}\gdu{b}\t^\b\brsb\nn\\
    \S^{abc}&=&d\blsb-\frac{35}{3}x^a dx^b dx^c -3x^{[a}dx^b\gdu{c]}-\quart x^{[a}\gdu{b}\gdu{c]}\brsb\nn.
\eea
\subsection{$p=4$ charges}
\bea
    \S^{\a\b\g\d}&=&d\blsb d\t^\a d\t^\b d\t^\g \t^\d\brsb\\
    \S^{a\b\g\d}&=&d\blsb 6dx^a d\t^\b d\t^\g \t^\d+\frac{3}{5}\gdu{a}d\t^{(\b}d\t^{\g}\t^{\d)}\brsb\nn\\
    \S^{ab\b\g}&=&d\blsb-19dx^a dx^b d\t^\b \t^\g-3dx^{[a}\gdu{b]}d\t^{(\b}\t^{\g)}+x^{[a}\gdu{b]}d\t^\b d\t^\g\nn\\
        &&-\frac{1}{4}\gdu{a}\gdu{b}d\t^{(\b}\t^{\g)}\brsb\nn\\
    \S^{abc\b}&=&d\blsb-\frac{65}{2}dx^a dx^b dx^c \t^\b-\frac{19}{4}dx^{[a}dx^b\gdu{c]}\t^\b+6x^{[a}dx^b\gdu{c]}d\t^\b\nn\\
        &&-\frac{7}{8}dx^{[a}\gdu{b}\gdu{c]}\t^\b+\half x^{[a}\gdu{b}\gdu{c]}d\t^\b\nn\\
        &&-16\gdu{a}\gdu{b}\gdu{c}\t^\b\brsb\nn\\
    \S^{abcd}&=&d\blsb -21x^a dx^b dx^c dx^d -\frac{19}{2}x^{[a}dx^b dx^c \gdu{d]}-\frac{7}{4}x^{[a}dx^b\gdu{c}\gdu{d]}\nn\\
        &&-\frac{1}{8}x^{[a}\gdu{b}\gdu{c}\gdu{d]}\brsb\nn.
\eea
\subsection{$p=5$ charges}
\bea
\label{3:p=5 representation}
    \S^{\a\b\g\d\e}&=&d\blsb d\t^\a d\t^\b d\t^\g d\t^\d \t^\e\brsb\\
    \S^{a\b\g\d\e}&=&d\blsb\frac{15}{2}dx^a d\t^\b d\t^\g d\t^\d \t^\e+\frac{5}{6}\gdu{a}d\t^{(\b}d\t^{\g}d\t^{\d}\t^{\e)}\brsb\nn\\
    \S^{ab\b\g\d}&=&d\blsb-\frac{47}{2}dx^a dx^b d\t^\b d\t^\g \t^\d-\frac{9}{2}dx^{[a}\gdu{b]}d\t^{(\b}d\t^{\g}\t^{\d)}\nn\\
        &&-x^{[a}\gdu{b]}d\t^\b d\t^\d\t^\d-\frac{3}{8}\gdu{a}\gdu{b}d\t^{(\b}d\t^{\g}\t^{\d)}\brsb\nn\\
    \S^{abc\b\g}&=&d\blsb-52dx^a dx^b dx^c d\t^\b \t^\g-\frac{47}{4}dx^{[a}dx^b\gdu{c]}d\t^{(\b}\t^{\g)}\nn\\
        &&-\frac{15}{2}x^{[a}dx^b\gdu{c]}d\t^\b\d\t^\g-\frac{17}{8}dx^{[a}\gdu{b}\gdu{c]}d\t^{(\b}\t^{\g)}\nn\\
        &&-\frac{5}{8}x^{[a}\gdu{b}\gdu{c]}d\t^\b d\t^\g-\frac{7}{48}\gdu{a}\gdu{b}\gdu{c}d\t^{(\b}\t^{\g)}\brsb\nn\\
    \S^{abcd\b}&=&d\blsb\frac{281}{4}dx^a dx^b dx^c dx^d \t^\b+13dx^{[a}dx^b dx^c \gdu{d]}\t^\b\nn\\
        &&+\frac{47}{2}x^{[a}dx^b dx^c\gdu{d]}d\t^\b+\frac{31}{8}dx^{[a}dx^b\gdu{c}\gdu{d]}\t^\b\nn\\
        &&+\frac{17}{4}x^{[a}dx^b\gdu{c}\gdu{d]}d\t^\b+\frac{7}{12}dx^{[a}\gdu{b}\gdu{c}\gdu{d]}\t^\b\nn\\
        &&+\frac{7}{24}x^{[a}\gdu{b}\gdu{c}\gdu{d]}d\t^\b+\frac{7}{192}\gdu{a}\gdu{b}\gdu{c}\gdu{d}\t^\b\brsb\nn\\
    \S^{abcde}&=&d\blsb\frac{77}{2}x^a dx^b dx^c dx^d dx^e+26x^{[a}dx^{b}dx^c dx^d \gdu{e]}\nn\\
        &&+\frac{31}{4}x^{[a}dx^b dx^c \gdu{d}\gdu{e]}+\frac{7}{6}x^{[a}dx^b \gdu{c}\gdu{d}\gdu{e]}\nn\\
        &&+\frac{7}{96}x^{[a}\gdu{b} \gdu{c}\gdu{d}\gdu{e]}\brsb\nn.
\eea

Having found a representation of the $\S^{A_1\ldots A_p}$, we now need to check the validity of the ansatze (\ref{4:p=1 anomalous term}) and (\ref{p greater than 1 anomalous term}) for the corresponding anomalous term representatives. Firstly, one verifies using the relevant Fierz identity that $sM^{(p)}=0$. $M^{(p)}$ is also identically $d$ closed since the $\S^{A_1\ldots A_p}$ are closed forms. We therefore have $M^{(p)}\in H^{p,2}$. Because [M] is the unique, $D$ nontrivial class, \textit{any} nontrivial representative of $H^{p,2}$ is a representative of $[M]$. It therefore suffices to show that $M^{(p)}$ is $D$ nontrivial. The coboundaries of $H^{p,2}$ are identically equal to the gauge transformations. Hence, if there exists a gauge field $\lambda\in\Omega^{p-1,1}$ such that:
\be
\label{trivial M}
    M^{(p)}=sd\lambda
\ee
then $M^{(p)}$ is trivial (since then $M^{(p)}=Dd\lambda$). Otherwise it is nontrivial.

In the case of the superstring, it was explicitly shown that $M^{(1)}$ is $D$ cohomologous to $H$ \cite{reimers05}. The nontriviality of $M^{(1)}$ then follows from that of $H$. For $p\geq 2$, one notes that $M^{(p)}$ is constructed using the structure constants $\G_{a_1...a_p\a\b}$, $\G_{a\a\b}$ and $\eta_{ab}$. In attempting to solve (\ref{trivial M}), one therefore needs to consider only those $\lambda$ gauge fields constructed using these constants. We believe the following to be a complete set of such fields:
\bea
\label{set of lambda fields}
    \lambda^{(i)}&=&x^{a}dx^{a_1}...dx^{a_i}\gdu{a_{i+1}}...\gdu{a_{p-1}}\bar e \G_{aa_1...a_{p-1}}\t,\\
        &&\qquad 0\leq i\leq p-1.\nn\\
    \lambda'^{(i)}&=&\bar e \G^{a}\t x^{b}dx^{a_1}...dx^{a_i}\gdu{a_{i+1}}...\gdu{a_{p-2}}\bar \t \G_{aba_1...a_{p-2}}d\t,\nn\\
        &&\qquad 0\leq i\leq p-2\nn.\\
    \lambda''^{(i)}&=&e^{a}x^{b}dx^{a_1}...dx^{a_i}\gdu{a_{i+1}}...\gdu{a_{p-2}}\bar \t \G_{aba_1...a_{p-2}}d\t,\nn\\
        &&\qquad 0\leq i\leq p-2\nn.
\eea
In equation (\ref{trivial M}), it suffices to consider the terms of highest order in $x^m$. One then needs to consider a linear combination of only $\lambda^{(p-1)}$ and $\lambda''^{(p-2)}$. One finds that a solution for the coefficients does not exist for any value of $p$. Provided that the set (\ref{set of lambda fields}) is complete, $M^{(p)}$ is therefore nontrivial, and is thus a representative for the anomalous term associated with the standard superspace $p$-brane action. The charges (\ref{3:p=1 representation}) through (\ref{3:p=5 representation}) (and their associated anomalous terms) generalize the results of \cite{azc89} to the case where fermionic topological terms are retained. Note that for $p\geq 3$ there are additional charges not present in the anomalous term; these are simply those which close the extended algebra (they result from the action of $Q_A$ on the anomalous term). We conclude that the algebras (\ref{p=1 algebra}) through (\ref{p=5 algebra}) are indeed generated as topological charge algebras of the standard $p$-brane action.

\section{Extended superspace actions}
\label{sec:extended Noether charges}

The extended superalgebras (\ref{p=1 algebra}) through (\ref{p=5 algebra}) can be used to construct left invariant potentials $B$ for the field strength $H$ \cite{siegel94,bergshoeff95,chrys99}. The corresponding extended superspace $p$-brane action is the same as (\ref{action}), but where $B$ is now the left invariant potential. In this case, $W=0$ solves the descent equations, and the corresponding Noether charge algebra is the minimal algebra. In \cite{chrys99}, Noether charges associated with the extra coordinates of the $p=1,2$ extended superspace actions were found. Equations (\ref{3:p=1 representation}, \ref{3:p=2 representation}) are proportional to the forms given there. Although these results were obtained in different contexts\footnote{In the previous section we constructed topological charges of the standard superspace action and showed that they generate the extended algebras (\ref{p=1 algebra}) through (\ref{p=5 algebra}). We may contrast this with the work of \cite{chrys99}, where the extended algebras were used from the outset to construct invariant extended superspace actions. The resulting Noether charges associated with extra coordinates were then found for the cases $p=1,2$. It was noted there that the bosonic topological term of these charges agrees with that obtained from the anomalous term of the standard superspace formulation \cite{azc89}. Showing that this correspondence also holds for the fermionic topological terms is a new result which is the main purpose of this section.}, they should agree. In each case the forms transform according to the same extended superalgebra, and we claim that based upon this transformation property alone the solution is unique.

Conversely, it follows that our results extend those of \cite{chrys99} to give the Noether charges associated with the extra coordinates of the extended superspace actions for the remaining values of $p$. Let us separate the generators of the underlying extended superalgebras into standard/extended parts as $T_{A}=\{T_{\widetilde A},T_{\check A}\}$, with:
\bea
    T_{\widetilde A}&=&\lcb -\widetilde Q_\a,-\widetilde P_a\rcb \\
    T_{\check A}&=&\lcb -\S_{\check A}\rcb \nn\\
    &=&\lcb -\S^{A_1\ldots A_p}\rcb \nn.
\eea
The extra generators $T_{\check A}$ form an ideal. It follows that the standard coordinates do not transform under the left/right group actions generated by $T_{\check A}$. The inverse vielbeins therefore satisfy:
\bea
    R_{\check A}{}^{\widetilde M}&=&0\\
    L_{\check A}{}^{\widetilde M}&=&0\nn.
\eea
Now, the momenta of the action can be written \cite{azc91,reimers05}:
\be
    P_{M}=P^{(NG)}_{M}+(i_{\del_1}\ldots i_{\del_p}B)_{M},
\ee
where $i$ is the interior derivation and $\del_i$ is the $i$-th worldvolume tangent vector. $P^{(NG)}_{M}$ are the conjugate momenta for the NG action, which vanish for the extra coordinates:
\be
    P^{(NG)}_{\check M}=0.
\ee
It follows that for the extended superspace action, the Noether charge associated with the generator $T_{\check A}$ is that derived (slightly differently) in \cite{chrys99}:
\bea
    \label{extended Noether charge}
    {\bar Q}_{\check A}&=&\int d^p\s R_{\check A}{}^{M}(i_{\del_1}\ldots i_{\del_p}B)_{M}\\
    &=&\bar {(i_{V_{\check A}}B)}\nn,
\eea
where:
\be
    V_{\check A}=R_{\check A}{}^{M}\del_{M}
\ee
is the left invariant vector field associated with $T_{\check A}$. Since the Noether charges satisfy the extended superalgebras (\ref{p=1 algebra}) through (\ref{p=5 algebra}) under Poisson brackets, it follows that the forms $i_{V_{\check A}}B$ must satisfy the same algebra under the action of $Q_{\widetilde A}$. We claim that forms satisfying this transformation property have the unique solutions (\ref{3:p=1 representation}) through (\ref{3:p=5 representation}). So, for an appropriate normalization of the action one has:
\be
    i_{V_{\check A}}B=\S_{\check A},
\ee
and the Noether charges:
\be
    \label{equality of charges}
    {\bar Q}_{\check A}=\bar\S_{\check A}.
\ee
Interestingly enough, this argument has explicitly determined some Noether charges for a $p$-brane action without needing the explicit structure of the WZ term. It is only required that the extended background superspace must \textit{admit} a left invariant WZ form. That such WZ forms do indeed exist was shown explicitly for $p\leq 3$ by constructing the required potential $B$ \cite{bergshoeff95,chrys99}.

The conserved charges $\S^{A_1\ldots A_p}$ are thus the same in both the standard and extended superspace formulations of the action. In the former they are anomalous terms of the Noether charge algebra, while in the latter they are the Noether charges themselves. This result extends that of \cite{chrys99} to establish correspondence between fermionic as well as bosonic terms, and also for all allowed values of $p$.

\section{Comments}
\label{sec:Comments}

The representations for $\S^{A_1\ldots A_p}$ appear to be a basis for the $p$-forms. It seems possible to invert each representation to write:
\be
    dx^{m_1}\ldots dx^{m_i}d\t^{\m_1}\ldots d\t^{\m_{p-i}}\leftrightarrow \{ \S^{a_1\ldots a_j\a_1\ldots \a_{p-j}},\quad j\leq i\}.
\ee
For example, for $p=2$:
\bea
    d\t^\a d\t^\b&=&\S^{\a\b}\\
    dx^a d\t^\a&=&\frac{2}{9}\S^{a\a}+\frac{1}{18}\t^\a \G^a{}_{\b\g}\S^{\b\g}-\frac{1}{18}\gu{a}{\b}\S^{\a\b}\nn\\
    dx^a dx^b&=&-\frac{1}{5}\S^{ab}+\frac{1}{45}\gu{[a}{\a}\S^{b]\a}-\frac{1}{10}x^{[a}\G^{b]}{}_{\a\b}\S^{\a\b}\nn\\
    &&-\frac{1}{180}\gu{a}{\a}\gu{b}{\b}\S^{\a\b}\nn.
\eea
This constitutes a change of basis for the $p$-forms, which in this case is not inherited in the usual way from a vielbein.

The topological anomalous term $\bar M^{(p)}$ is a topological integral of its form representation $M^{(p)}$. If the fermionic topology is taken to be trivial, then the only contribution to $\bar M^{(p)}$ comes from the $(dx)^p$ term of $\S^{a_1\ldots a_p}$. This is the ``central" anomalous term found in \cite{azc89}. The corresponding extended algebra can be related to partial breaking of supersymmetry \cite{sorokin97,townsend97}. We note that this central extension is not present in all gauges. Using the gauge transformation generated by $\lambda^{(p-1)}$ one finds that the fully modified Noether charge algebra in the presence of trivial fermionic topology is:
\be
    \left \{\widetilde {\bar Q}_\a,\widetilde {\bar Q}_\b\right \}=-\G^a{}_{\a\b}\widetilde {\bar P}_a-E\G_{a_1\ldots a_p\a\b}\int dx^{a_1}\ldots dx^{a_p},
\ee
where the integral is over the spatial section of the brane, and $E$ is a free constant resulting from the $\lambda$ gauge freedom. The familiar bosonic extension of the $p$-brane Noether charge algebra is thus the result of a specific choice of gauge. In another gauge one obtains the minimal algebra\footnote{A free multiplicative constant also results from an optional tension parameter normalizing the action \cite{azc89}. In this case one obtains the minimal algebra only in the limiting case of zero tension (the action used here vanishes at the limit). Tension and gauge parameters have completely different effects when fermionic topological terms are retained; in that case there may be multiple anomalous terms and the gauge parameters are not global scale factors.}.

A precursor to the $p=2$ algebra (\ref{p=2 algebra}) was an algebra that results from setting $\S^{\a\b}=0$ in (\ref{p=2 algebra}) \cite{Bergshoeff89}. This algebra does not appear in the spectrum of topological charge algebras generated by the standard action. One may see this by noting that $\S^{a\b}$ becomes ``central" in this algebra. Since the only left invariant possibilities for a form representing this generator are not closed, this cannot be a topological charge algebra. This might also have been expected on the basis that this contracted algebra does not allow the construction of a left invariant WZ form \cite{bergshoeff95} (topological charge algebras of the standard action appear to be such that they \textit{do} allow the construction of such WZ forms \cite{reimers05}). Although $\S^{\a\b}$ appears to be a necessary generator in topological charge algebras, it's possible for the associated \textit{anomalous term} $M_{ab}$ to vanish (and commuting translations are thus restored: $[P_a,P_b]=0$). For example, to obtain such algebras for $p=2$, one first applies the gauge transformation generated by $\half\lambda''^{(0)}$ from (\ref{set of lambda fields}), which sets $M_{ab}=0$. All remaining gauge transformations then preserve this property.

One may ask if there are any new algebras of interest generated as topological charge algebras of the standard action. Upon investigating the set of $p=2$ gauge transformations (\ref{set of lambda fields}) we found that new superalgebras were generated which allowed the construction of left invariant WZ forms. However, they seem to require the introduction of more generators than are present in (\ref{p=2 algebra}). Upon constructing the left invariant WZ form, one then finds that free parameters remain. This is because the space has been extended more than is necessary; one might say that the associated superspace is not ``minimally extended." In the case $p=1$, we found that the entire spectrum of topological charge algebras yielded minimally extended superspaces \cite{reimers05}. However, for $p\geq 2$ it appears that (\ref{p=2 algebra}) through (\ref{p=5 algebra}) may be the unique, minimally extended topological charge algebras generated by the standard $p$-brane action.

\subsection{Acknowledgments}
I would like to thank I. N. McArthur for helpful suggestions and critical reading of the manuscript.

\bibliographystyle{hieeetr}
\bibliography{double_complex}
\end{document}